\documentstyle[12pt]{article}
\newcommand{\nut}{
\begin{picture}(11,5)(0,0)
\put(5,-10){$\tilde{}$}
\put(0,0){\it N}
\end{picture}}

\newcommand{\mut}{
\begin{picture}(12,5)(0,0)
\put(6,-10){$\tilde{}$}
\put(0,0){\it M}
\end{picture}}

\newcommand{\eut}{
\begin{picture}(3,3)(-2,-2)
\put(-2,-15){$\tilde{}$}
\put(-5,-2){$\eta$}
\end{picture}}

\begin{document}

\hspace{11.6cm} CGPG-94/10-3

\begin{center}

\baselineskip=24pt plus 0.2pt minus 0.2pt
\lineskip=22pt plus 0.2pt minus 0.2pt

 \Large

Real Ashtekar Variables for Lorentzian Signature Space-times\\

\vspace*{0.35in}

\large

J.\ Fernando\ Barbero\ G. $^{\ast, \dag}$
\vspace*{0.25in}

\normalsize

$^{\ast}$Center for Gravitational Physics and Geometry,\\
Department of Physics,\\
Pennsylvania State University,\\
University Park, PA 16802\\
U.S.A.\\

$^{\dag}$Instituto de Matem\'aticas y F\'{\i}sica Fundamental, \\
C.S.I.C.\\
Serrano 119--123, 28006 Madrid, Spain
\\

\vspace{.5in}
October 10, 1994\\
\vspace{.5in}

ABSTRACT

\end{center}
\vspace{.5in}

I suggest in this letter a new strategy to attack the problem of the
reality conditions in the Ashtekar approach to classical and quantum
general relativity. By writing a modified Hamiltonian constraint in the
usual $SO(3)$ Yang-Mills phase space I show that it is possible to
describe space-times with Lorentzian signature without the introduction of
complex variables. All the features of the Ashtekar formalism related to
the geometrical nature of the new variables are retained; in particular,
it is still possible, in principle, to use the loop variables approach in
the passage to the quantum theory.  The key issue in the new formulation
is how to deal with the more complicated Hamiltonian constraint that must
be used in order to avoid the introduction of complex fields.

\pagebreak

\baselineskip=24pt plus 0.2pt minus 0.2pt
\lineskip=22pt plus 0.2pt minus 0.2pt

\setcounter{page}{1}

The purpose of this letter is to suggest a new strategy to deal with the
problem of the reality conditions in the Ashtekar approach to classical
and quantum gravity. At the present moment there is some consensus about
the reasons behind the success of the Ashtekar variables program
\cite{Asht}. One of them is the geometrical nature of the new variables.
In particular, the fact that the configuration variable is a connection is
specially interesting because this allows us to use loop variables both at
the classical and quantum level \cite{Smo}. Another advantage of the
formalism is the simplicity of the constraints --specially the Hamiltonian
constraint-- that has been very helpful in finding solutions to all of
them. There are, however, some difficulties in the formalism that must be
solved and are not present in the traditional ADM scheme \cite{ADM}. The
most conspicuous one is the fact that complex variables must be used in
order to describe Lorentzian signature space-times. This is often put in
relation with the fact that the definition of self-duality in these
space-times demands the introduction of imaginary coefficients. The now
accepted way to deal with this issue is the introduction of reality
conditions. They impose some consistency requirements on the scalar
product in the Hilbert space of physical states. In fact, the hope is that
this scalar product can be selected by the reality conditions. There are,
however some difficulties with this approach too. Specifically it is very
difficult to implement the reality conditions in the loop variables
scheme. Only recently some positive results in this direction have been
reported \cite{Jos}. The main point of this letter is to consider the
geometrical nature of the Ashtekar variables as the most important asset
of the formalism. With this idea in mind, it is easy to see that the
introduction of complex variables is necessary only if one wants to have
an specially simple form for the Hamiltonian constraint. If we accept to
live with a more complicated Hamiltonian constraint in the Ashtekar phase
space we can use real variables.

An interesting consequence of this, as emphasized by Rovelli and Smolin,
is that all the results obtained within the loop variables approach
(existence of volume and area observables, weave states and so on
\cite{SSS}, \cite{ARS}) whose derivation is independent of the particular
form of the scalar constraint scalar can be maintained even for Lorentzian
signature space-times because it is possible to describe Lorentzian
gravity with real fields in the Ashtekar phase space. More specifically,
the issue is not the implementation of the reality conditions (at least at
the kinematical) but rather the construction of a scalar product,
normalizability of the quantum physical states and so on. The proposal
presented in this letter does not address this problems. It must also be
said that the construction of area and volume observables referred to
above must still be put in a completely sound and rigorous mathematical
basis that may very well be provided by the approach presented in
\cite{Jos} to incorporate the reality conditions in the loop variables
approach by using a generalization of th e Bargmann-Siegel transform to
spaces of connections. This letter has nothing to add with respect to
this.  In the following, tangent space indices and $SO(3)$ indices will be
represented by lowercase Latin letters from the beginning and the middle
of the alphabet respectively. The 3-dimensional Levi-Civita tensor density
and its inverse will be denoted\footnote{I represent the density weights
by the usual convention of using tildes above and below the fields.} by
$\tilde{\eta}^{abc}$ and $\;\eut_{abc}$ and the internal $SO(3)$
Levi-Civita tensor by $\epsilon_{ijk}$. The variables in the $SO(3)$-ADM
phase space (ADM formalism with internal $SO(3)$ symmetry as discussed in
\cite{ABJ}) are a densitized triad $\tilde{E}_{i}^{a}$ (with determinant
denoted by $\tilde{\!\tilde{E}}$) and its canonically conjugate object
$K_{a}^{i}$ (closely related to the extrinsic curvature). The (densitized)
three dimensional metric built from the triad will be denoted
$\tilde{\;\;\tilde{q}^{ab}}\equiv \tilde{E}^{a}_{i}\tilde{E}^{bi}$ and its
determinant $\tilde{\!\tilde{q}}$ so that
$q^{ab}=\frac{\tilde{\;\;\;\tilde{q}^{ab}}}{ \;\tilde{\!\tilde{q}}}$.  I
will use also the $SO(3)$ connection $\Gamma_{a}^{i}$ compatible with the
triad.  The variables in the Ashtekar phase space are $\tilde{E}_{i}^{a}$,
again, and the $SO(3)$ connection $A_{a}^{i}$. The curvatures of
$A_{a}^{i}$ and $\Gamma_{a}^{i}$ are respectively given by
$F_{ab}^{i}\equiv 2\partial_{[a} A_{b]}^{i}+\epsilon^{i}_{\;\;jk}A_{a}^{j}
A_{b}^{k}$ and $R_{ab}^{i}\equiv 2\partial_{[a}
\Gamma_{b]}^{i}+\epsilon^{i}_{\;\;jk}\Gamma_{a}^{j} \Gamma_{b}^{k}$.
Finally, the action of the covariant derivatives defined by these
connections on internal indices are\footnote {They may be extended to act
on tangent indices, if necessary, by introducing a space-time torsion-free
connection; for example the Christoffel connection $\Gamma_{ab}^{c}$ built
from $q^{ab}$. All the results presented in the paper will be independent
of such extension}
$\nabla_{a}\lambda_{i}=\partial_{a}\lambda_{i}+\epsilon_{ijk}A_{a}^{j}
\lambda^{k}$
and ${\cal
D}_{a}\lambda_{i}=\partial_{a}\lambda_{i}+\epsilon_{ijk}\Gamma_{a}^{j}
\lambda^{k}$.
The compatibility of $\Gamma_{a}^{i}$ and $\tilde {E}_{i}^{a}$
thus means ${\cal
D}_{a}\tilde{E}_{i}^{b}\equiv\partial_{a}\tilde{E}^{b}_{i}+
\epsilon_{i}^{\;\;jk}\Gamma_{aj}
\tilde{E}^{b}_{k}+\Gamma_{ac}^{b}\tilde{E}^{c}_{i}-\Gamma_{ac}^{c}
\tilde{E}^{b}_{i}=0$.

I will start from the $SO(3)$-ADM constraints
\begin{eqnarray}
& & \epsilon_{ijk}K_{a}^{j}\tilde{E}^{ak}=0\nonumber\\
& & {\cal D}_{a}\left[\tilde{E}^{a}_{k}K_{b}^{k}-\delta_{b}^{a}
\tilde{E}^{c}_{k}
K_{c}^{k}\right]=0\label{1}\\
& & -\zeta\sqrt{\tilde{\!\tilde{q}}}R+\frac{2}{\sqrt{\tilde{\!\tilde{q}}}}
\tilde{E}^{[c}_{k}\tilde{E}^{d]}_{l}K_{c}^{k}K_{d}^{l}=0\nonumber
\end{eqnarray}
where $R$ is the scalar curvature of the three-metric $q_{ab}$ (the inverse of
$q^{ab}$).
The variables $K_{ai}(x)$ and $\tilde{E}^{b}_{j}(y)$ are canonical;
i.e. they satisfy
\begin{eqnarray}
& & \left\{K_{a}^{i}(x), K_{b}^{j}(y)\right\}=0\nonumber\\
& & \left\{\tilde{E}^{a}_{i}(x),
 K_{b}^{j}(y)\right\}=\delta_{j}^{i}\delta_{a}^{b}\delta^{3}(x,y)
\label{111}\\
& &  \left\{\tilde{E}^{a}_{i}(x), \tilde{E}^{b}_{j}(y)\right\}=0\nonumber
\end{eqnarray}
The parameter $\zeta$ is used to control the space-time signature. For
Lorentzian signatures we have $\zeta=-1$ whereas in the Euclidean case we
have $\zeta=+1$. The constraints (\ref{1}) generate internal $SO(3)$
rotations, diffeomorphisms and time evolution. I write now the usual
canonical transformation to the Ashtekar phase space
\begin{eqnarray}
& & \tilde{E}^{a}_{i}= \tilde{E}^{a}_{i}\label{2}\\
& & A_{a}^{i}=\Gamma_{a}^{i}+\beta K_{a}^{i}\label{3}
\end{eqnarray}
here $\beta$ is a free parameter that I will adjust later. The Poisson brackets
between the new variables $A_{a}^{i}$ and $\tilde{E}^{a}_{i}$ are
\begin{eqnarray}
& & \left\{A_{a}^{i}(x), A_{b}^{j}(y)\right\}=0\nonumber\\
& & \left\{A_{a}^{i}(x),
\tilde{E}^{b}_{j}(y)\right\}=-\beta\delta_{j}^{i}\delta_{a}^{b}\delta^{3}(x,y)
\label{4}\\
& &  \left\{\tilde{E}^{a}_{i}(x), \tilde{E}^{b}_{j}(y)\right\}=0\nonumber
\end{eqnarray}
and thus, the transformation is canonical. Introducing (\ref{2}, \ref{3}) in
the constraints (\ref{1}) we get immediately the following constraints in the
Ashtekar
phase space
\begin{eqnarray}
& & \tilde{G}_{i} \equiv \nabla_{a}\tilde{E}^{a}_{i}=0\label{5}\\
& & \tilde{V}_{a} \equiv F_{ab}^{i}\tilde{E}^{b}_{i}=0\label{6}\\
& & \tilde{\!\tilde{S}} \equiv -\zeta
\epsilon^{ijk}\tilde{E}^{a}_{i}\tilde{E}^{b}_{j}F_{abk}+\frac{2(\beta^{2}
\zeta-1)}
{\beta^{2}}\tilde{E}^{a}_{[i}\tilde{E}^{b}_{j]}(A_{a}^{i}-\Gamma_{a}^{i})
(A_{b}^{j}-
\Gamma_{b}^{j})=0\label{7}
\end{eqnarray}
They are the Gauss law, vector and scalar constraints of the Ashtekar
formulation. The traditional attitude with regard to (\ref{7}) has been to
consider that the last term introduces unnecessary complications in the
formalism.  For this reason it has always been cancelled by choosing
$\beta$ such that $\beta^{2}\zeta-1=0$. For Euclidean signatures we can
take $\beta^{2}=1$ and remain within the limits of the real theory. For
Lorentzian signatures, however, we are forced to take $\beta^{2}=-1$ and
then the variables (specifically the connection) cease to be real. It must
be emphasized that this is true only if we insist in cancelling the
last term in
(\ref{7}). If we accept to keep it, there is no reason to introduce complex
objects in the theory. The value of $\beta$ (as long as it is different
from zero) is
also irrelevant so we can choose $\beta=-1$ and have the following Hamiltonian
constraint in the Lorentzian case
\begin{equation}
\epsilon^{ijk}\tilde{E}^{a}_{i}\tilde{E}^{b}_{j}F_{abk}-4
\tilde{E}^{a}_{[i}\tilde{E}^{b}_{j]}(A_{a}^{i}-\Gamma_{a}^{i})(A_{b}^{j}-
\Gamma_{b}^{j})=0\label{8}
\end{equation}
The relevant Poisson bracket in (\ref{4}) becomes
\begin{equation}
\left\{A_{a}^{i}(x),
\tilde{E}^{b}_{j}(y)\right\}=\delta_{j}^{i}\delta_{a}^{b}\delta^{3}(x,y)
\label{88}
\end{equation}
Since we have obtained this result by performing a canonical transformation,
the Poisson
algebra of the constraints is preserved. If we define the functionals
\begin{eqnarray}
& & G[N^{i}]\equiv \int d^{3}x\;N^{i}\tilde{G}_{i}\nonumber\\
& & V[N^{a}]\equiv\int d^{3}x\;N^{a}\tilde{V}_{a}\label{9}\\
& & S[\nut]\equiv\int d^3x\;\nut\tilde{\!\tilde{S}}\nonumber
\end{eqnarray}
we have the usual Poisson algebra; in particular the Poisson bracket of
$S[\nut]$ and
$S[\mut]$ is given by
\begin{equation}
\left\{S[\nut],
S[\mut]\right\}=+V[\tilde{E}^{a}_{i}\tilde{E}^{bi}(\nut\partial_{b}\mut-
\mut\partial_{b}\nut)]\label{10}
\end{equation}
The $+$ sign in the right-hand side of (\ref{10}) shows that we have, indeed,
Lorentzian signature. It is possible to rewrite (\ref{8}) in a more
appealing form.
The second term, in particular, can be expressed in terms of covariant
derivatives of
$\tilde{E}^{a}_{i}$.
To this end I introduce the inverse of
$\frac{\;\tilde{E}^{a}_{i}}{\sqrt{\tilde{\!\tilde{E}}}}$
\begin{equation}
e_{ai}\equiv\frac{1}{2\sqrt{\tilde{\!\tilde{E}}}}\;\;\eut_{abc}\epsilon^{ijk}
\tilde{E}^{b}_{j}\tilde{E}^{c}_{k}\label{11}
\end{equation}
where $\tilde{\!\tilde{E}}\equiv \det \tilde{E}^{a}_{i}$ and rewrite
(\ref{8}) in the form
\begin{eqnarray}
\hspace{-1.5cm}\epsilon^{ijk}\tilde{E}^{a}_{i}\tilde{E}^{b}_{j}F_{abk}-
\tilde{\eta}^{a_{1}a_{2}a_{3}}\tilde{\eta}^{b_{1}b_{2}b_{3}}
\left[(e_{a_{1}}^{i}
\nabla_{a_{2}}e_{a_{3}i}) (e_{b_{1}}^{j}\nabla_{b_{2}}e_{b_{3}j})\right. & &
\label{12} \\
& \left.\hspace{-2.5cm}-2(e_{a_{1}}^{j}\nabla_{a_{2}}e_{a_{3}i})
(e_{b_{1}}^{i}\nabla_{b_{2}}e_{b_{3}j})\right]=0 &
\nonumber
\end{eqnarray}
The last term in the previous formula is still quadratic in the
connections but its dependence on $\tilde{E}^{a}_{i}$ is complicated. It
must be noted also that if we restrict ourselves to non-degenerate triads
it can be cast in polynomial form (of degree 8 in $\tilde{E}^{a}_{i}$) by
multiplying it by the square of $\tilde{\!\tilde{E}}$.  If one is
interested in checking explicitly the Poisson algebra and use the
Hamiltonian constraint (\ref{12}), it is useful to notice that
$e_{ai}[\tilde{E}]$ and $ -2\tilde{\eta}^{abc}\nabla_{b}e_{ci}$ are
canonically conjugate objects. This may eventually be useful in order to
write the new Hamiltonian constraint in terms of loop variables maybe by
allowing us to extend the set of T-variables with objects built out of
$e_{ai}[\tilde{E}]$ and $-2\tilde{\eta}^{abc}\nabla_{b}e_{ci}$.

There is another appealing way to write a Hamiltonian constraint for
Lorentzian general relativity in terms of real Ashtekar variables. One
starts by writing the Hamiltonian constraint in the $SO(3)$-ADM formalism
in the form
\begin{eqnarray}
-2\zeta\sqrt{\tilde{\!\tilde{q}}}R+\zeta\sqrt{\tilde{\!\tilde{q}}}R+
\frac{2}{\sqrt{\tilde{\!\tilde{q}}}}\tilde{E}^{[c}_{k}\tilde{E}^{d]}_{l}
K_{c}^{k}K_{d}^{l}=\hspace{3cm} & & \label{13}\\
-2\zeta\sqrt{\tilde{\!\tilde{q}}}R-\frac{1}{\sqrt{\tilde{\!\tilde{q}}}}
\left[\zeta
\epsilon^{ijk}\tilde{E}^{a}_{i}\tilde{E}^{b}_{j}F_{abk}-\frac{2(\beta^{2}
\zeta+1)}
{\beta^{2}}\tilde{E}^{a}_{[i}\tilde{E}^{b}_{j]}(A_{a}^{i}-\Gamma_{a}^{i})
(A_{b}^{j}-
\Gamma_{b}^{j})\right]=0\nonumber
\end{eqnarray}
Now, in the Lorentzian case we can choose $\beta^{2}=1$ and cancel the
last term to give
\begin{equation}
2\;\tilde{\!\tilde{q}}R+
\epsilon^{ijk}\tilde{E}^{a}_{i}\tilde{E}^{b}_{j}F_{abk}=0\label{14}
\end{equation}
remembering now that
\begin{equation}
\tilde{\!\tilde{q}}R=-\epsilon^{ijk}\tilde{E}^{a}_{i}\tilde{E}^{b}_{j}R_{abk}
\label{15}
\end{equation}
we can finally write the Hamiltonian constraint as\footnote{It is my
understanding
that this formulation was independently considered by Ashtekar \cite{ooo}
before the loop variables formalism had been introduced, and
discarded due to the presence of the potential term.}
\begin{equation}
\epsilon^{ijk}\tilde{E}^{a}_{i}\tilde{E}^{b}_{j}(F_{abk}-2R_{abk})=0
\label{16}
\end{equation}
The geometrical interpretation of the term that we must add to the
familiar Hamiltonian constraint in the Ashtekar formulation in order to
describe Lorentzian gravity in the Ashtekar phase space is simpler than in
(\ref{12}); it is just the curvature of the $SO(3)$ connection compatible
with the triad $\tilde{E}^{a}_{i}$. Some comments are now in order.

 First, the presence of a potential term in (\ref{12}) and (\ref{16})
certainly makes them more complicated than the familiar Ashtekar
Hamiltonian constraint. Taking into account that one of the sources of
difficulties in the ADM formalism is precisely the presence of a potential
term in the Hamiltonian constraint (see \cite{KU} and references therein
for examples on how the quantization of ADM gravity would simplify in the
absence of such a term) it is fair to expect some difficulties in the
treatment of the theory with this new Hamiltonian constraint. The
simplification brought about by removing the reality conditions has been
traded for a more complicated Hamiltonian constraint.

The way the difference between the Euclidean and Lorentzian cases arises
is rather interesting; there is potential term in the Lorentzian case that
is absent in the Euclidean formulation. This asymmetry between the
Euclidean and Lorentzian cases is somehow puzzling. Why is it that the
"complicated formulation" is found for the Lorentzian case? In the ADM
formalism such an asymmetry is not apparent in the formalism.

The fact that the theory is written in an $SO(3)$ Yang-Mills phase space
makes it possible to attempt its quantization by using loop variables.
This can be achieved in principle because we know \cite{Jur} that loop
variables are good coordinates (modulo sets of measure zero) in the (Gauss
law reduced) constraint hypersurface. The key problem is now how to write
the potential term in terms of the familiar loop variables. The obvious
solution would be to add additional objects built with traces of
holonomies of the connection $\Gamma_{a}^{i}$, notice, however, that it is
not straightforward to add them to the set of elementary variables $T^{0}$
and $T^{1}$ because this would spoil the closure under the Poisson
brackets. It is worthwhile noting that the possibility of writing the
Hamiltonian constraint for real Lorentzian general relativity in the two
alternative forms (\ref{12}) and (\ref{16}) may be useful when trying to
write them in terms of loop variables. It is conceivable that one form may
be simpler to deal with than the other.

The form of the constraints of the theory makes it possible to use an
approach similar to that of Capovilla, Dell and Jacobson in \cite{CDJ} to
solve both the vector and scalar constraints. We define, for non
degenerate triads, the matrix $\psi_{ij}$ as
\begin{equation}
\psi_{ij}\tilde{E}^{a}_{j}=\tilde{B}^{a}_{i}-2\tilde{R}^{a}_{i}
\label{17}
\end{equation}
where
\begin{eqnarray}
& & \tilde{B}^{c}_{i}\equiv \tilde{\eta}^{abc}F_{abi}\label{18}\\
& &  \tilde{R}^{c}_{i}\equiv \tilde{\eta}^{abc}R_{abi}\label{18a}
\end{eqnarray}
the scalar constraint is then
\begin{equation}
\epsilon^{ijk}\;\eut_{abc}\tilde{E}^{a}_{i}\tilde{E}^{b}_{j}\psi_{kl}
\tilde{E}^{c}_{l}=
2\;\tilde{\!\tilde{E}}tr\psi=0\Longrightarrow\;\; tr \psi=0
\label{19}
\end{equation}
The vector constraint can be rewritten now as
\begin{equation}
\tilde{E}^{a}_{i}(F_{ab}^{i}-2R_{ab}^{i})=0\Longleftrightarrow
\tilde{E}^{[a}_{i}(\tilde{B}^{b]i}-2\tilde{R}^{b]i})=0
\label{20}
\end{equation}
because the relation
$R_{ab}^{i}=-\frac{1}{2}\epsilon^{ijk}R_{abc}^{\;\;\;\;\;\;d}e_{c}^{j}
e_{d}^{k}$ and
the Bianchi identity $R_{[abc]}^{\;\;\;\;\;\;d}=0$ (the three
dimensional Riemann tensor built with $q^{ab}$) imply that
$\tilde{E}^{a}_{i}R_{ab}^{i}=0$. We have then
\begin{equation}
\tilde{E}^{[a}_{i}\tilde{E}^{b]}_{j} \psi_{ij}=0\Longrightarrow\;\;
\psi_{[ij]}=0
\label{21}
\end{equation}
so that a symmetric and traceless $\psi_{ij}$ solves both the vector and
scalar constraints. As in the usual case we are left with one last
equation: the Gauss law. Here is where the main difference between the
usual Hamiltonian constraint and (\ref{16}) arises. Without the potential
term of (\ref{16}) we could very easily write the remaining equation in
terms of $A_{ai}$ and $\psi_{ij}$
\begin{equation}
\nabla_{a}[\psi^{-1}_{ij}\tilde{B}^{a}_{j}]=0
\label{22}
\end{equation}
Now the situation is more complicated because we are forced to consider
a system of coupled PDE's
\begin{eqnarray}
& & \nabla_{a}[\psi^{-1}_{ij}(\tilde{B}^{a}_{j}-2\tilde{R}^{a}_{j})]=0
\label{23}\\
& & \psi_{ij}\tilde{E}^{a}_{j}=\tilde{B}^{a}_{i}-2\tilde{R}^{a}_{i}\label{24}
\end{eqnarray}
The second equation could be solved, in principle, for $\tilde{E}^{a}_{i}$
and then
the first would become an equation for $\psi_{ij}$ and $A_{a}^{i}$ only as in
(\ref{22}).

The main result presented in this letter has been the introduction of
several alternative forms for the Hamiltonian constraint for Lorentzian
space-times in the Ashtekar formalism with real variables. The problem of
implementing the reality conditions in the theory has been transformed
into the problem of working with the new Hamiltonian constraints
introduced here.

The previous results strongly suggest that Lorentzian general relativity
is a theory of two $SO(3)$ connections (in the sense that both the
curvatures of $A_{a}^{i}$ and $\Gamma_{a}^{i}$ seem to be playing a role
as is apparent in (\ref{16})). A completely different two-connection
formulation for both Euclidean and Lorentzian general relativity has been
reported elsewhere \cite{Fer}. In that formulation the main difference
between the Euclidean and Lorentzian cases is the appearance of terms
depending on the difference of the curvatures for the Lorentzian signature
case. The fact that, even for Lorentzian signatures, the Hamiltonian
constraint of that formulation is a low order polynomial of the curvatures
makes it suitable to be written in terms of loop variables built with the
two connections. My hope is that the comparison of the several different
approaches discussed above may provide useful information about the way to
proceed with the quantization program for general relativity and the role
of complex fields in it.

{\bf Acknowledgements} I wish to thank A. Ashtekar, P. Peld\'an and L.
Smolin for their remarks and comments and the Spanish Research Council
(CSIC) for providing financial support.

\end{document}